\documentclass[aps,prx,twocolumn,final,letterpaper,floatfix]{revtex4-2}

\usepackage{appendix}
\usepackage{graphicx}   
\usepackage{import}
\usepackage{comment}
\usepackage{epstopdf}
\usepackage{bm}
\usepackage{amssymb}
\usepackage{quotes}
\usepackage{braket}
\usepackage{transparent}
\usepackage{dcolumn}
\usepackage{multirow}
\usepackage{cancel} 
\usepackage{mdframed}
\usepackage{color}
\usepackage{bm}
\usepackage{dsfont}
\usepackage{slashed}
\usepackage{enumitem}
\usepackage{amsmath}

\begin{document}
\rmfamily

\title{Long-range entanglement and quantum correlations in a multi-frequency comb system}

\author{Sahil Pontula$^{1,2,3}$}
\email{spontula@mit.edu}
\author{Debasmita Banerjee$^4$}
\author{Marin Solja\v{c}i\'{c}$^{1,3}$}
\author{Yannick Salamin$^{4}$}
\email{yannick.salamin@ucf.edu}

\affiliation{
$^1$Department of Physics, MIT, Cambridge, MA 02139, USA. \\
$^2$Department of Electrical Engineering and Computer Science, MIT, Cambridge, MA 02139, USA.\\
$^3$Research Laboratory of Electronics, MIT, Cambridge, MA 02139,  USA.\\
$^4$CREOL, The College of Optics and Photonics, University of Central Florida, Orlando, Florida 32816, USA.
}

\begin{abstract}

Frequency combs are multimode photonic systems that underlie countless precision sensing and metrology applications. Since their invention over two decades ago, numerous efforts have pushed frequency combs to broader bandwidths and more stable operation. More recently, quantum squeezing and entanglement have been explored in single frequency comb systems for quantum advantages in sensing and signal multiplexing. However, the production of quantum light across multiple frequency combs remains unexplored.
In this work, we theoretically explore a mechanism that generates a series of nonlinearly coupled frequency combs through cascaded three-wave upconversion and downconversion processes mediated by a single idler comb. We show how this system generates inter- and intracomb two-mode squeezing and entanglement spanning a very large spectral range, from ultraviolet to mid-IR frequencies. Finally, we show how this system can be engineered to produce on-demand multimode quantum light through covariance matrix optimization. Our findings could enable tunable broadband ``ghost'' spectroscopy protocols, squeezing-enhanced pump-probe measurements, and broadband entanglement between spectrally-multiplexed quanta of information.   

\end{abstract}
\maketitle


\section{\label{sec:level1}Introduction}

Optical frequency combs have revolutionized precision metrology, providing a coherent link between the optical and microwave domains while enabling advances in spectroscopy, timing, and sensing \cite{udem2002optical, diddams2010evolving}. In parallel, nonlinear optical interactions have become the subject of growing interest for generating quantum light in multimode optical systems \cite{presutti2024highly, guidry2022quantum}. For example, strong quantum correlations in multimode systems can manifest as noise reduction below the standard quantum limit in appropriate ``supermodes'' and quadrature combinations \cite{guidry2023multimode, roslund2014wavelength}, offering pathways for quantum-enhanced sensing protocols.
Therefore, continuous-variable entanglement and squeezing in frequency comb systems have attracted significant attention due to their potential for quantum-enhanced measurements and quantum information processing  \cite{roslund2014wavelength, herman2025squeezed, shi2023entanglement, hariri2025entangled}. 
Quantum comb line correlations and continuous variable (CV) multipartite entanglement have been demonstrated in both bulk (e.g., synchronously pumped optical parametric oscillators) \cite{pinel2012generation, roslund2014wavelength, medeiros2014full, chen2014experimental} and integrated (e.g., microresonator) systems \cite{wang2025large, yang2021squeezed, jia2025continuous, jahanbozorgi2022continuous, shen2025highly, jahanbozorgi2023generation, li2023multicolor, chembo2016quantum, guidry2022quantum, bensemhoun2024multipartite, reimer2016generation}. Furthermore, multipartite entanglement and cluster states in frequency combs and optical parametric oscillators \cite{wang2025large, chen2014experimental, barbosa2018hexapartite, pysher2011parallel, gerke2015full, roslund2014wavelength, villar2006direct, cai2017multimode} show rapid progress towards large, scalable frequency-multiplexed quantum resources for computation, networks, and metrology. Beyond single combs, two-comb systems have found key applications in dual comb spectroscopy (DCS) and have recently been extended to the quantum domain \cite{herman2025squeezed, coddington2016dual, shi2023entanglement, wan2025quantum, hariri2025entangled}. However, quantum correlations and entanglement remain restricted to a single comb and limited spectral range.


To address this gap, we investigate the quantum optics of multi-comb systems by examining continuous-variable entanglement and quantum correlations in a nonlinear system that generates a comb of above-threshold frequency combs. Such nested comb structures introduce additional degrees of freedom and coupling pathways that fundamentally alter the multimode quantum landscape compared to single- and two-comb systems. 
We demonstrate that the nonlinear processes that couple different combs create long-range multipartite entanglement and quantum correlations that span both the intra-comb and inter-comb domains. Intriguingly, these quantum behaviors extend to modes that do not couple directly via the nonlinear mechanism. These quantum correlations manifest as complex squeezing patterns in collective quadratures that encompass modes from multiple combs. We emphasize that, in contrast to the considerable work devoted to producing single, ultra-broadband frequency combs \cite{picque2019frequency, timmers2018molecular, zhang2019broadband, luke2015broadband, hu2022high, adler2010mid, hoghooghi2022broadband}, we are specifically interested in the classical and quantum behavior of a system of distinct, coupled frequency combs that can potentially span a very large spectral range.

The physical underpinnings of our multi-comb system lie in cascaded three-wave mixing processes (TWMs). The TWMs are mediated by a common low-frequency idler comb (initially generated by parametric downconversion of two optical combs) that propagates in the system and distributes energy into upconverted and downconverted combs. In addition to exploring entangled frequency comb generation, we show how our proposed physical mechanism addresses two major gaps in the field of quantum frequency combs: (1) long-range entanglement between pairs of modes in vastly different spectral ranges, which is conventionally limited by the comb span, and (2) programmable multimode quantum light generation through covariance matrix optimization, which we achieve through dissipation and pump engineering.    

The rich quantum resource presented by this multi-comb architecture opens new possibilities for quantum-enhanced applications. For example, we discuss how the hierarchical entanglement structure we explore could enable novel forms of entanglement-assisted spectroscopy using multiple bright combs, as well as spectral multiplexing in CV quantum computing protocols. Furthermore, noting that most frequency combs operate well above the shot noise limit, our results suggest how multimode states with strong quantum correlations can be impactful for low-noise applications. Our findings establish a framework for understanding and harnessing quantum correlations in these complex optical systems, paving the way for new quantum technologies based on hierarchical multi-frequency comb architectures.

\begin{figure*}
    \centering
    \includegraphics[scale=0.65]{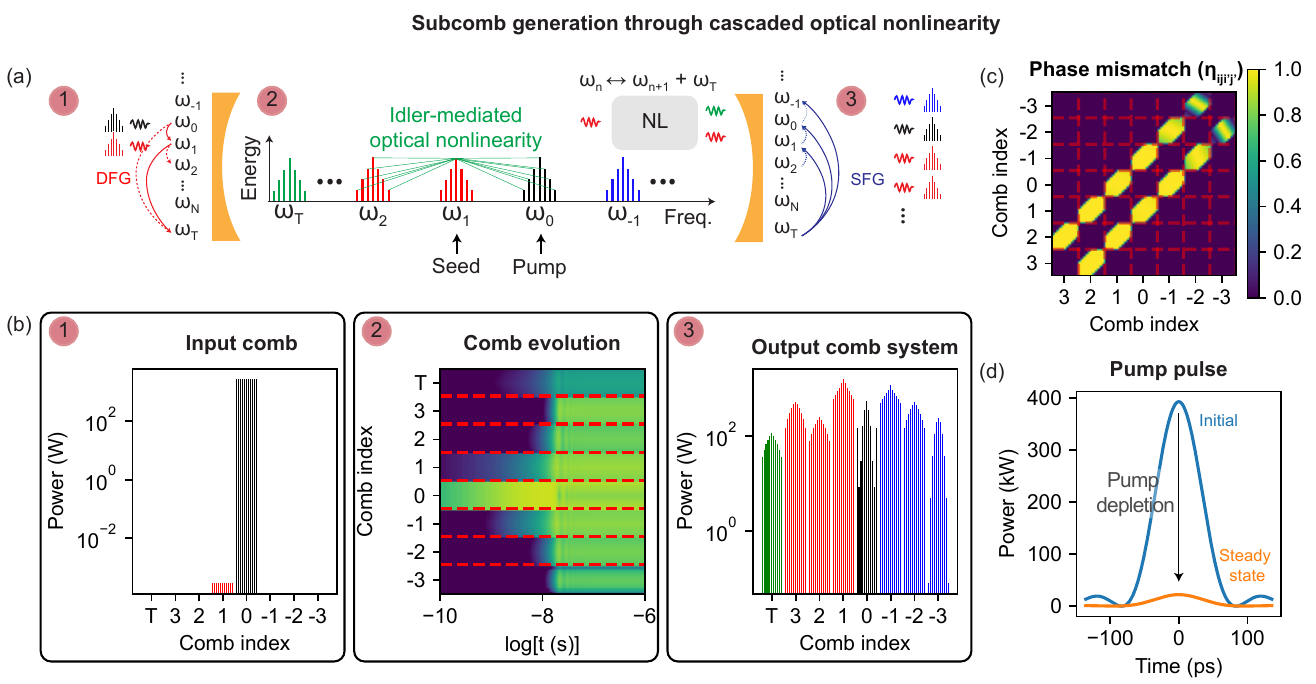}
    \caption{\textbf{Cascaded comb generation in a multimode cavity.} (a) A multimode system is pumped with a comb centered at frequency $\omega_0$ and seeded with either a single frequency or comb centered at frequency $\omega_1$. DFG between the pump and seed generates an idler comb at frequency $\omega_T$, which is then recycled to generated ``subcombs'' through cascaded sum and difference frequency generation (SFG/DFG) processes. Taken together, the subcombs comprise a ``primary comb'' with frequency spacing $\omega_T$. This system can be realized in a multimode cavity that is synchronously pumped/seeded by a high repetition rate laser, which gives rise to the subcomb frequency spacing $\omega_m\ll\omega_T$. Cascaded nonlinear processes produce the primary comb with spacing $\omega_T$ centered around the pump subcomb at frequency $\omega_0$. (b) Cascaded comb generation in a multimode cavity with $I=7$ subcombs, each with $J=11$ modes. (c) Heatmap of dispersion-induced phase matching factor between individual modes  $\eta_{iji'j'}=|\mathrm{sinc}(\Delta k_{iji'j'}L)|$ ($\Delta k_{iji'j'}=k_{ij}-k_{i'j'}-(\omega_{ij}-\omega_{i'j'})/c$ is the wavevector mismatch, $L$ is the crystal length, and lithium niobate dispersion is assumed \cite{zelmon1997infrared}). (d) Depletion of the pump pulse due to the cascaded nonlinear interactions. All comb modes have quality factor $Q_0=5\times 10^6$, the idler comb has quality factor $Q_T=10^4$, and $\beta_0=3\times 10^{-4}$ J$^{-1/2}$. The pump wavelength is $\lambda_{0}=465$ nm, the idler wavelength is $\lambda_T=4.07$ $\mu$m ($\omega_T=2\pi\cdot 73.7$ THz), and the intracomb spacing is $\omega_m=2\pi\cdot 1$ GHz. The subcombs lie at wavelengths 708, 603, 525, 465, 417, 378, and 346 nm. In (b), ``T'' denotes the idler comb.}
    \label{fig:1}
\end{figure*}

\section{Cascaded comb generation in a multimode nonlinear cavity}

The physical mechanism we explore is depicted in Fig. \ref{fig:1}, which features cascaded comb generation in a multimode nonlinear cavity. In the most general case, a pump and seed comb centered at frequencies $\omega_0\equiv \omega_{00}$ and $\omega_1\equiv \omega_{10}$ are injected into the cavity and undergo three-wave mixing (TWM) to generate an idler comb centered at $\omega_T=\omega_0-\omega_1$. The idler comb is then recycled in the cavity to generate blueshifted ($\omega_{n<0}$) and redshifted ($\omega_{n>0}$) combs relative to $\omega_0$ through nonlinear upconversion and downconversion processes \cite{pontula2024shaping, pontula2025non}. 


Fig. \ref{fig:1} features the cascaded generation of ``subcombs'' through sum/difference frequency generation processes (SFG/DFG). The subcombs collectively comprise a ``primary comb'' which has spacing $\omega_T\equiv \omega_{T0}$ equal to the center frequency of the common idler comb that mediates the cascaded nonlinear processes. The system is pumped with a comb centered at frequency $\omega_0$ with comb line spacing $\omega_m\ll \omega_T$ (which can come from synchronized pumping of a multimode cavity by a high repetition rate or electro-optically modulated laser). We note that the condition $\omega_m\ll\omega_T$ can be satisfied for a wide variety of systems across the electromagnetic spectrum, as we discuss below. Because the idler comb couples all of the $\omega_m$-spaced subcombs, one may anticipate nontrivial correlations and entanglement to emerge between discrete modes of different subcombs.  
Writing the Hamiltonian for this nonlinear multimode system in the discrete frequency mode basis, we obtain the following Heisenberg equations of motion in the mean field that describe the system dynamics:
\begin{align}
    \begin{split}
        \dot a^T_{-k} = &\sum_{ij}\beta_{ijk}^{-} a^*_{ij}a_{i-1,j-k} - (\gamma_T + \mu_T) a^T_{-k} \\
        \dot a_{ij} = &\sum_k \left[-\beta^{+*}_{ijk} a^T_{-k} a_{i+1,j+k} + \beta^-_{ijk} a^{T*}_{-k} a_{i-1,j-k} \right] \\&- (\gamma_{ij} + \mu_{ij}) a_{ij} + \sqrt{2\gamma_{ij}}s_{ij}.
    \end{split}
    \label{eq:subcomb}
\end{align}
Here, $a^T_{-k}$ denotes the mode in the idler comb at frequency $\omega_{T0}+k\omega_m$ and $a_{ij}$ indexes mode $j$ in subcomb $i$ with frequency $\omega_{ij}=\omega_{00}-i\omega_{T0}-j\omega_m$. As described in the SI, $\beta^{\pm}_{ijk}=\beta_0\eta_{i,j,i\pm 1, j\pm k}\sqrt{\hbar\omega_{ij}\omega_{i\pm 1,j\pm k} \omega^T_{-k}}$ denotes the nonlinear strength, with $\beta_0$ related to the nonlinear susceptibility $\chi^{(2)}$ of the cavity's nonlinear crystal that supports the cascaded SFG/DFG processes. The phase matching prefactor $\eta_{iji'j'}=|\mathrm{sinc}(\Delta k_{iji'j'}L)|$, where $\Delta k_{iji'j'}=k_{ij}-k_{i'j'}-(\omega_{ij}-\omega_{i'j'})/c$ is the wavevector mismatch and $L$ is the crystal length, governs the strength of nonlinear coupling between individual frequency modes. $\gamma_{ij},\mu_{ij}$ respectively denote outcoupling and intrinsic loss for each comb mode (while $\gamma_T, \mu_T$ denote the corresponding losses for the idler comb, assumed constant across the comb for simplicity) and the $s_{ij}$ indicate which modes in the system are externally pumped. Here, we consider $s_{i\notin \{0,1\},j}=0$. 

In Fig. \ref{fig:1}b, we explore how the multimode nonlinear system takes the energy in the initial pump comb(s) and uses it to create combs at different frequencies, both blueshifted and redshifted relative to the pump. Three-wave mixing interactions between adjacent combs populate the idler comb (green). The strength of mixing between individual frequency modes is governed by phase matching, as shown in Fig. \ref{fig:1}c. This phase matching heatmap determines which modes mix within the phase matching bandwidth and therefore determines the extent of the idler comb and other subcombs (for small $\omega_m$ or thin crystals, the phase matching bandwidth is larger and more modes between adjacent combs are allowed to mix with one another). Changing $\omega_m$ is one simple form of dispersion engineering that can significantly impact both the mean field and quantum noise behavior of the system, as we explore in the SI. Finally, in Fig. \ref{fig:1}d, we show how the cascaded comb generation is accompanied by significant depletion of the pump, as energy is being moved into the cascading subcombs. It is in this pump depletion regime where we expect to see signatures of strong quantum correlations and entanglement.

\section{Two-mode squeezing and quantum correlations}

By linearizing the full Heisenberg-Langevin equations of motion about the steady state obtained by solving Eq. \ref{eq:subcomb} (see SI for details), we calculate the intracavity fluctuations $\delta \hat p_{k,\mathrm{in}}=\delta \hat a_k + \delta \hat a_k^\dagger$ (amplitude/sum quadrature) and $\delta \hat q_{k,\mathrm{in}}=-i(\delta \hat a_k - \delta \hat a_k^\dagger)$ (phase/difference quadrature) for each mode, as described in more detail in the SI. The output fluctuation (annihilation) field operators $\delta \hat s_{k,\mathrm{out}}=\sqrt{2\gamma_k}\delta \hat a_k-\delta \hat s_{k,\mathrm{in}}$ can then be computed from the input-output relation of temporal coupled mode theory \cite{joannopoulos2008molding}, as described in the SI. We can define the output amplitude operator $\delta p_{k,\mathrm{out}}=\delta \hat s_{k_\mathrm{out}} + \delta \hat s_{k,\mathrm{out}}^\dagger$ and phase operator $\delta q_{k,\mathrm{out}}$ (defined similarly). Unless otherwise specified, we will use the shorthand $\delta\hat p_k=\delta p_{k,\mathrm{out}}$ and similarly for $\delta\hat q_k$. Output intensity fluctuations can be derived from the fluctuations of $\hat P_k = \hat s_{k,\mathrm{out}}^\dagger \hat s_{k,\mathrm{out}}$.


In Fig. \ref{fig:2}, we explore the mean field and quantum noise behavior of a system that in steady state possesses $I=5$ subcombs, each with $J=3$ comb lines. Fig. \ref{fig:2}a presents a schematic illustration of one possible platform for this system, where an electro-optic modulated quasi-continuous wave source pumps a nonlinear multimode cavity. The modulation rate sets the spacing $\omega_m$ of modes within each subcomb. The modulator can also be placed inside the cavity to increase the subcomb length \cite{parriaux2020electro}. In this case, $\omega_m$ should be chosen to correspond to a multiple of the free spectral range, so that all comb modes are resonant in the cavity. The interplay between the modulation (which introduces linear coupling between the modes in individual subcombs) and the optical nonlinearity (which couples subcombs) may lead to interesting mean field and quantum noise behaviors \cite{pontula2025non}. Finally, the intracavity nonlinear crystal is phase matched for cascaded SFG/DFG. Note that, if the pump and seed are very nondegenerate (as they are here), a composite crystal with multiple poling periods or multiple crystals are necessary to ensure phase matching for each three-wave mixing process (e.g., between combs 1 and 2, combs 0 and 1, etc.), as described in the SI.  

Fig. \ref{fig:2}b shows the individual energy and output intensity noise of each comb line. Note the pump depletion at $\omega_{00}$ and the fact that all modes are individually well above the shot noise limit because of the simultaneous noise coupling between modes from SFG and DFG processes. As we will discuss in Sec. \ref{sec:shaping} and as shown in the SI, a strong seed is a form of ``pump engineering'' that can yield shot noise limited comb lines and induce the formation of multimode low noise states.

Fig. \ref{fig:2}c explores multimode quantum states in the system. The cascading nonlinear processes introduce strong quantum correlations in excess of 20 dB below the uncorrelated limit that span the entire multi-comb system. This is evident in the heatmap of twin beam intensity difference noise, which plots $\langle (\delta \hat P_i - \delta\hat P_j)^2\rangle / \langle \delta \hat P_i^2 + \delta \hat P_j^2\rangle$ for all possible pairs of modes $\{\hat a_i,\hat a_j\}$. Importantly, because the combs can lie at very different frequencies, our mechanism can introduce strong quantum correlations across very different spectral ranges (for example, spanning from mid-IR to UV here). Furthermore, these low noise two-mode states also occur between combs that do not directly couple through the nearest-neighbor nonlinear interaction, suggesting that the idler comb is mediating quantum correlations spanning the entire multi-comb system.   

\section{Signatures of multipartite entanglement}

\begin{figure*}
    \centering
    \includegraphics[scale=0.65]{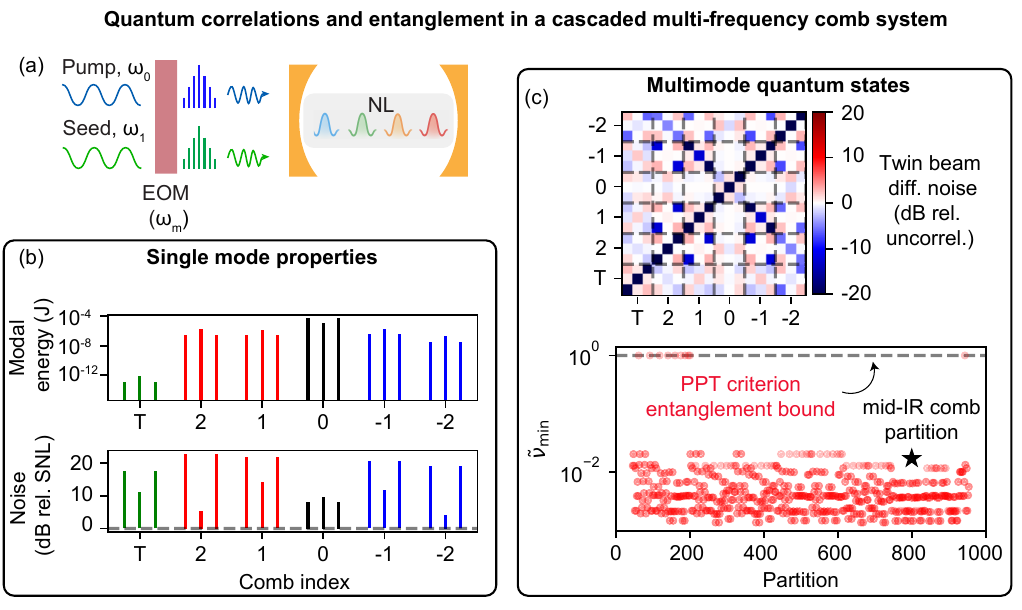}
    \caption{\textbf{Quantum correlations and entanglement in a multi-frequency comb system generated by cascaded second-order nonlinear processes.} (a) Schematic illustration of cascaded comb generation in a nonlinear (NL) multimode cavity pumped by an electro-optic modulated (EOM) quasi-continuous wave source. (b) Energy and output intensity noise (relative to shot noise limit) for individual comb modes. (c) Two-mode intensity difference noise and entanglement of the first 1,000 bipartitions as measured by the positive partial transform (PPT) criterion \cite{simon2000peres}. $\tilde\nu_\mathrm{min}$ denotes the minimum symplectic eigenvalue. The bipartition between the mid-IR idler comb and the other frequency modes is denoted by a star.
    Here, $I=5$ subcombs are simulated, each with $J=3$ modes; all comb modes have quality factor $Q_0=5\times 10^6$, the idler comb has quality factor $Q_T=10^5$, and $\beta_0=3\times 10^{-4}$ J$^{-1/2}$. The pump wavelength is $\lambda_{0}=465$ nm, $\lambda_T=4.07$ $\mu$m. The pump and seed powers are $|s_{0j}|^2 = 10$ kW, $|s_{1j}|^2=1$ mW. The modulation frequency is $\omega_m=2\pi\cdot 100$ MHz. The subcombs lie at wavelengths 603, 525, 465, 417, and 378 nm.}
    \label{fig:2}
\end{figure*}

Multiple metrics have been put forth to assess continuous variable entanglement in multimode systems \cite{simon2000peres, adesso2007entanglement, van2003detecting, duan2000inseparability}. Here, we use the well-known logarithmic negativity or partial positive transform (PPT) criterion \cite{simon2000peres}, which assesses the entanglement of an arbitrary bipartition of a system of $N$ modes through violation of the Robertson-Schrodinger (RS) uncertainty principle. This criterion is a sufficient condition for entanglement between the two partitions. Equivalently, one can look at the symplectic eigenvalues $\tilde\nu$ of the partially transposed covariance matrix and show violation of the RS principle through the presence of an eigenvalue $\tilde\nu<1$ (see SI for more details). This is the approach we take in Fig. \ref{fig:2}c, where we plot the minimum symplectic eigenvalue $\tilde\nu_\mathrm{min}$ for 1,000 of the $2^{N-1}-1$ total possible bipartitions, where $N=(I+1)J=18$ here ($I$ total subcombs and the idler comb, each with $J$ modes). The overwhelming majority of bipartitions show strong violation of the PPT criterion entanglement bound, the strength of entanglement scaling with $1/\tilde\nu_\mathrm{min}$ \cite{vidal2002computable, adesso2007entanglement}.

Notably, the quantum correlations and entanglement in this system span several distinct spectral ranges, from mid-IR to UV. Our system enables one to multiplex information in different comb systems and entangle stored information in spectrally separated subsets of modes. For example, we observe strong entanglement in the bipartition with one partition consisting of the mid-IR idler comb, represented by a star in Fig. \ref{fig:2}c. This entanglement confirms that the idler comb is indeed mediating quantum correlation with all other frequency modes in the system. The entanglement between one mode or comb with the rest of the system suggests exciting applications of ``ghost'' spectroscopy, where, owing to quantum entanglement, readout of a spectroscopic signal can occur in a spectral range that has never interacted with the sample of interest. 

Furthermore, although we limit our attention here to bipartite entanglement, the presence of two frequency dimensions (the primary and secondary comb dimensions) strongly suggests not only multipartite entanglement, but the possibility of multi-dimensional entanglement unimpeded by the scaling restrictions of real physical dimensions \cite{wang2018multidimensional}.

\section{Engineering on demand multimodal quantum states}
\label{sec:shaping}

\begin{figure*}
    \centering
    \includegraphics[scale=0.65]{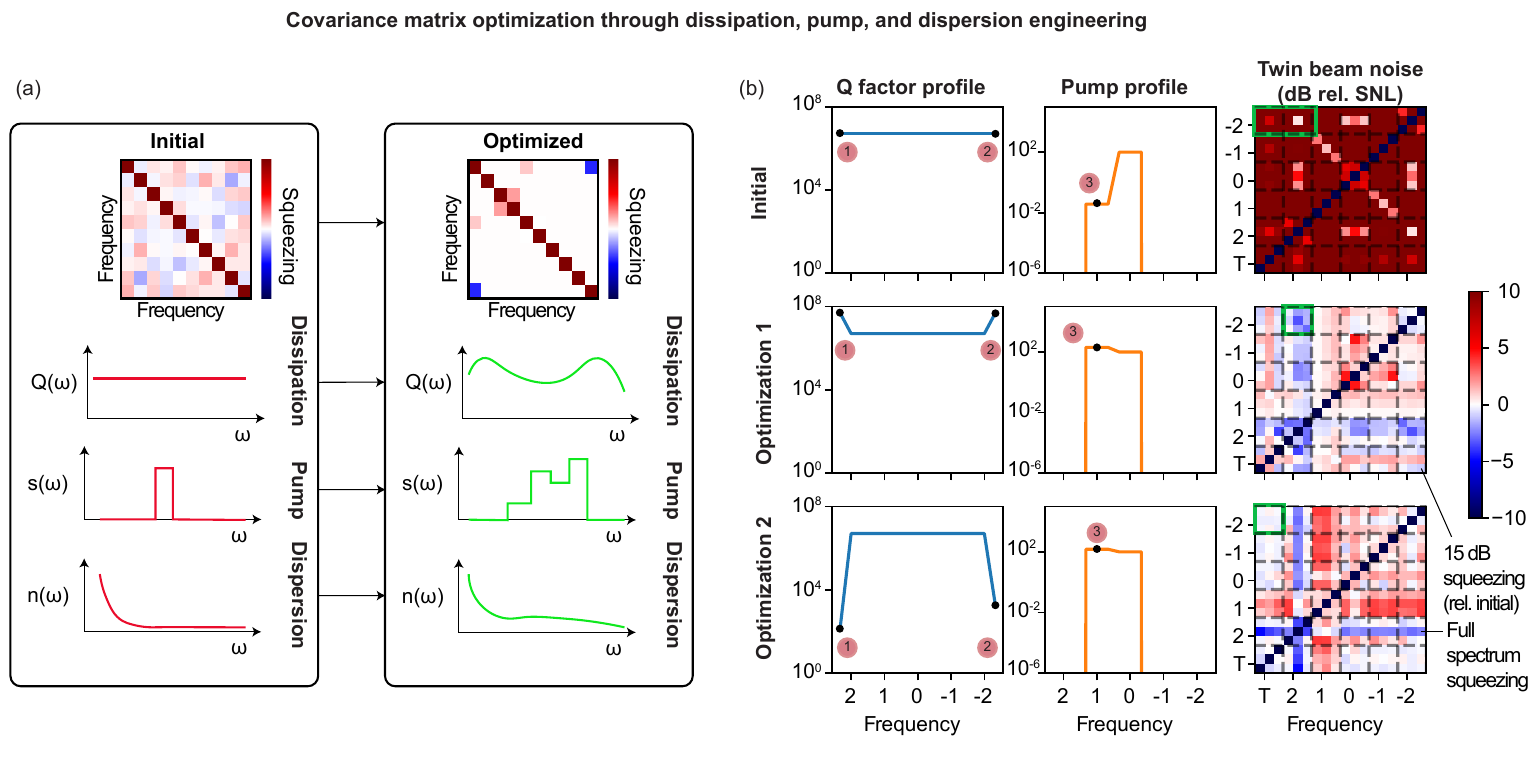}
    \caption{\textbf{Covariance optimization through dissipation, pump, and dispersion engineering.} (a) Illustrative optimization of dissipation via $Q$ factor (dissipation) spectrum $Q(\omega)$, pump spectrum $s(\omega)$, and dispersive refractive index $n(\omega)$ to achieve desired covariance matrix. (b) Two optimizations of the covariance matrix using a three parameter (endpoints of $Q$ factor spectrum and pump power of seed comb) constrained optimization. In Optimization 1, we maximize twin beam squeezing of the highest and lowest frequency modes, achieving 15 dB squeezing relative to the initial conditions. This optimization also achieves full comb-comb squeezing between the lowest and highest frequency optical combs, as shown by the green box. In Optimization 2, we maximize full comb-comb squeezing between the highest frequency optical comb and the mid-IR comb (green box), also achieving two-mode squeezing between a frequency mode in subcomb 2 and every other mode in the system (``full spectrum squeezing''). System parameters are identical to those in Fig. \ref{fig:2}. In (b), ``T'' denotes the idler comb.}
    \label{fig:4}
\end{figure*}

\begin{figure}
    \centering
    \includegraphics[scale=0.65]{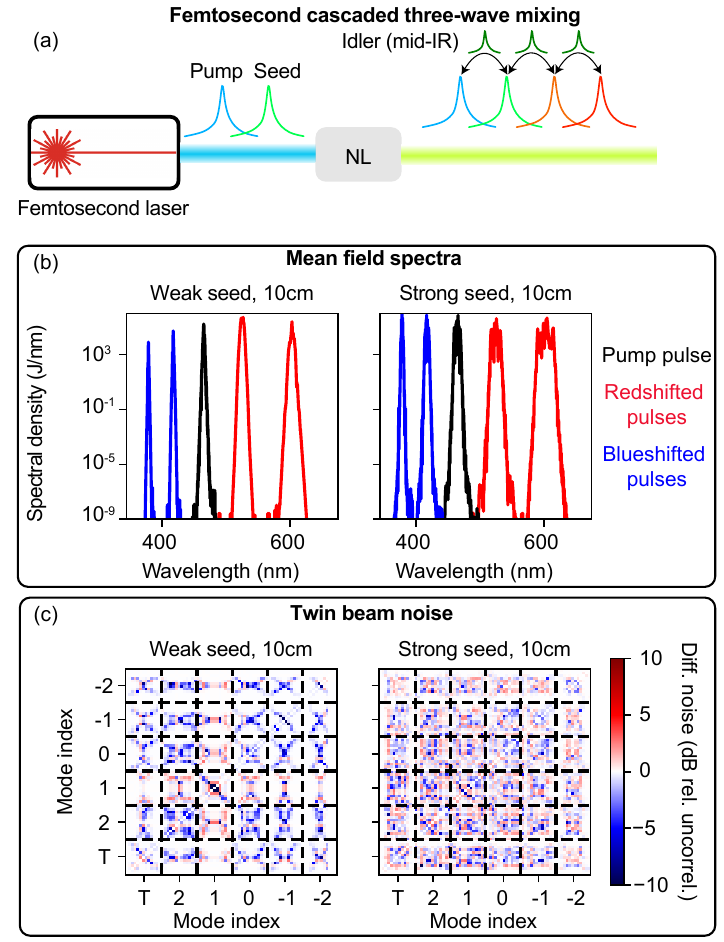}
    \caption{\textbf{Quantum correlations in a femtosecond pumped system with cascaded three-wave mixing processes.} (a) Schematic illustration of a femtosecond mode-locked laser pumping a $\chi^{(2)}$ nonlinear crystal (NL) phase matched for cascaded DFG and SFG. (b) Mean field spectra for the weak seed and strong seed limits. The idler pulse is not shown here. Spectra were calculated by integrating the system of nonlinear coupled generalized Schrodinger partial differential equations describing the equation of motion for each pulse. (c) Corresponding twin beam intensity difference noise extracted using quantum sensitivity analysis, where ``T'' denotes the idler pulse. Each square in the demarcated grid spans a range $\pm 32$ THz (resolution $0.52$ THz) around the center frequency of the mode. Pump wavelength: $\lambda_0=465$ nm, idler wavelength: $\lambda_T=4.07$ $\mu$m. Simulated nonlinear strengths are second-order nonlinearity $\beta^L\sim 6\times 10^{-17}$ W$^{-1/2}$m$^{-1}$s and third-order (Kerr) nonlinearity $\Gamma^L\sim 10^{-22}$ W$^{-1}$m$^{-1}$s. Dispersion was modeled through the refractive index profile of lithium niobate \cite{zelmon1997infrared}. Pump pulse parameters: average power 1 W, repetition rate 200 kHz, pulse duration 210 fs. Seed pulse parameters are identical, except for ``weak seed'' cases in which the average power is 10 mW. The propagation length in the nonlinear crystal is $L=10$ cm.}
    \label{fig:3}
\end{figure}

We next explore the possibility of on demand engineering of the covariance matrix, enabling the generation of arbitrary multimodal quantum states. This concept is illustrated schematically in Fig. \ref{fig:4}a, where we propose three tools for on demand optimization of the covariance matrix, namely dissipation, pump, and dispersion engineering. Dissipation engineering enables control over the loss ($Q$ factor) experienced by each frequency mode and the interplay between loss and nonlinear coupling enables one to shape nonlinear energy flow in the multi-comb system, affecting both mean field and quantum noise behavior. Through pump engineering, we can control the driving terms in the coupled mode equations. Finally, through dispersion engineering, one can tailor the nonlinear coefficient for each three-wave mixing process through tunable phase matching.

In Fig. \ref{fig:4}b, we explore two simple examples of covariance matrix optimization using the first two of the aforementioned tools (we explore the role of modified phase matching by changing $\omega_m$ in the SI). For simplicity, we only consider three optimization parameters, namely the $Q$ factors (parameters 1 and 2 in Fig. \ref{fig:4}b) of the first and last boundary modes in the subcomb system, each assumed to lie in the range $[10^2,10^8]$, and the pump power for the seed comb (parameter 3 in Fig. \ref{fig:4}b), assumed constant across the comb and constrained to the range $[10^{-3},10^4]$ W. More details on the optimization procedure are provided in the SI. Even with this highly constrained optimization, we show how tweaking only these three parameters allows significant reshaping of the covariance matrix. The initial covariance matrix with a weak seed and uniform $Q$ factor distribution across the subcombs features no two-mode low noise states, with most states well above 10 dB over the shot noise limit.

In our first optimization, we design our objective function to maximize the two-mode squeezing between the lowest and highest frequency modes in the system (here at mid-IR and UV frequencies). The optimization achieves sub-shot noise performance, nearly 15 dB below the initial twin beam noise between these two modes. Notice how the optimized covariance matrix simultaneously introduces several other low noise states between very disparate frequencies. For example, regions like those bounded by the green square feature low noise between entire combs (i.e. every comb line in one comb can form low noise states with every comb line in another comb).

In our second optimization, we design our objective function to achieve low noise states between the entire mid-IR idler comb and the highest frequency UV subcomb. Squeezing relative to the initial system exceeds 10 dB across both combs. Once again, the optimization achieves other low noise states, most notably the two-mode squeezing between a frequency mode in subcomb 2 and every other frequency mode in the system (``full spectrum squeezing'').

These results, while simple, showcase the promise of using dissipation and pump engineering for shaping the quantum noise and covariance matrix of multimode nonlinear systems. Although considerable work has explored the application of these and similar tools to the mean field distributions and classical dynamics of multimodal systems \cite{kim2017dispersion, fujii2020dispersion, hu2022mirror, chang2022integrated, song2024octave}, their application to generating tailored multimodal quantum states has remained until now unexplored. 

The forms of pump and dissipation engineering we consider here are readily achievable experimentally. The former is most directly controlled by simply changing the pump (or seed) power, but finer control may be offered by pulse shaping of the pump/seed \cite{weiner2011ultrafast}. The latter can be obtained from judicious choice of optical elements with frequency-dependent loss, such as photonic crystals, dichroic mirrors, or etalons/high-finesse Fabry-Perot resonators (for individual frequency mode control).  

\section{Broadband quantum correlations through cascaded ultrafast pulse generation}

\begin{figure*}
    \centering
    \includegraphics[scale=0.65]{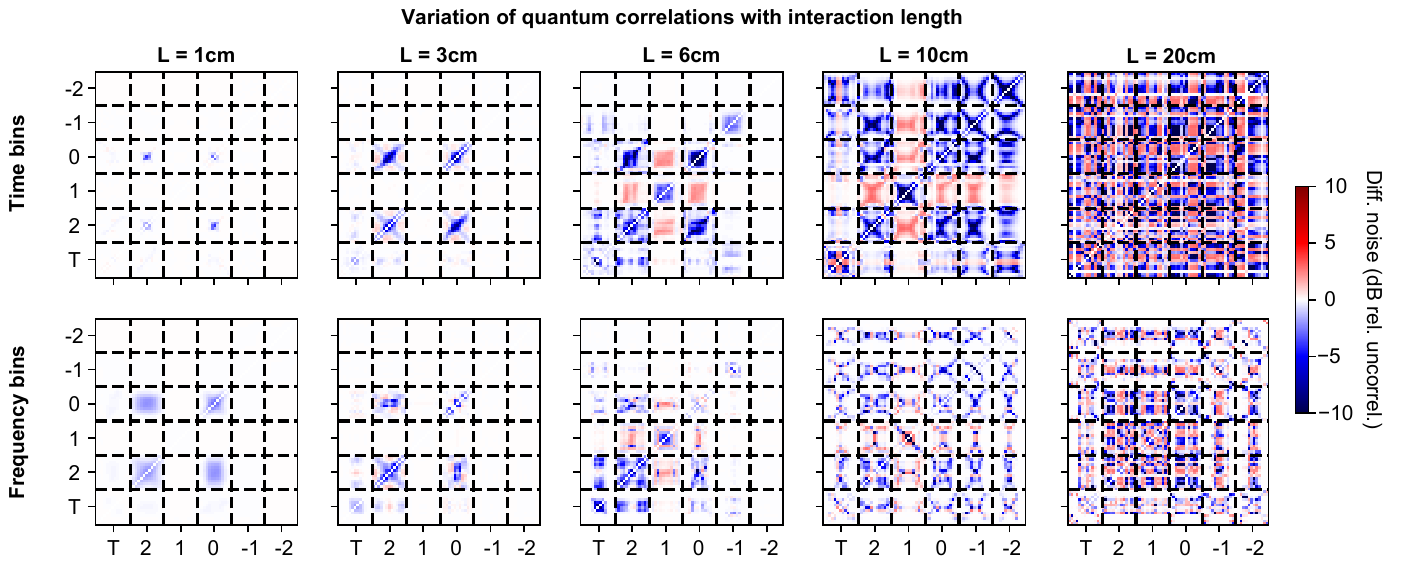}
    \caption{\textbf{Time-frequency quantum correlations in ultrafast cascaded three-wave mixing.} As the nonlinear interaction length increases, inter- and intrapulse quantum correlations begin to develop across the entire multi-pulse system in both the time and frequency domains. Simulation parameters are identical to those used in Fig. \ref{fig:3} for the ``weak seed'' case. Each dashed square spans a range $\pm 0.75$ ps (time bins, resolution 0.012 ps) or $\pm 32$ THz (frequency bins, resolution 0.52 THz) around the center of the pulse.}
    \label{fig:32}
\end{figure*}

Frequency combs are also commonly generated through femtosecond mode-locked lasers \cite{cundiff2003colloquium, diddams2007molecular, kim2016ultralow}, where the Fourier transform of a pulse train with repetition rate $f_m=\omega_m/2\pi$ corresponds to a frequency comb with spacing $f_m$. To complement our discrete mode analysis above, in this section, we calculate the quantum correlations in a multimode nonlinear system pumped by femtosecond pulses that supports cascaded SFG and DFG processes. Here, we simulate a single pulse propagating through a length $L$ of a nonlinear crystal supporting the cascaded second-order nonlinear processes that form the basis of our work (Fig. \ref{fig:3}a). In addition to second-order nonlinearity, we include the effects of Kerr nonlinearity, self-steepening, and dispersion (using the frequency-dependent refractive index of lithium niobate \cite{zelmon1997infrared}). The basis for our classical simulations is a system of coupled generalized nonlinear Schrodinger equations (GNLSEs), each equation describing the propagation of an individual pulse through the nonlinear medium. The GNLSE is a common method of simulating ultrafast pulse propagation in systems with dispersion and nonlinearity, such as optical fibers \cite{dudley2006supercontinuum, dudley2010supercontinuum, drummond2001quantum}. 

Our simulated system can represent several realistic experimental platforms, including both single pass (where each pump pulse interacts with the nonlinear crystal only once) and multi-pass configurations. The latter allows smaller nonlinear crystal lengths to be used while still retaining strong nonlinearity, and can take the form of a synchronously pumped optical parametric oscillator (SPOPO) \cite{cheung1990theory} or a configuration similar to a regenerative parametric amplifier \cite{reed1994widely}, where the pulses circulate in a (non-resonant) cavity multiple times before being outcoupled. 

In Fig. \ref{fig:3}b, we calculate the mean-field spectra for the system schematically shown in Fig. \ref{fig:3}a by propagating the coupled nonlinear system of $I+1$ GNLSEs forward for a length $L$ in the nonlinear crystal. For the pump pulse parameters we consider (chosen to approximate commercially available femtosecond laser systems \cite{mackevivciute2025comparative}), cascaded generation of pulses starts to occur appreciably at centimeter length scales, which generally lies in the pump depletion regime of operation. We consider two limits of operation: the weak seed limit ($|s_1|^2\ll |s_0|^2$) and the strong seed limit ($|s_1|^2\sim |s_0|^2$). Although the mean field spectra in these two cases are similar, their quantum noise properties are quite different as we will show.

To calculate quantum statistics in the ultrafast system we consider here, we use quantum sensitivity analysis (QSA), a recently developed method for efficiently calculating fluctuations in observables of a multimode nonlinear system from the Jacobian of a single classical simulation of the system's forward dynamics \cite{zia2025noise, rivera2025ultra}. As described in the SI, we calculate the covariance matrix of the system from the Jacobian, from which we plot the twin beam intensity difference noise as shown in Fig. \ref{fig:3}c.

Fig. \ref{fig:3}c presents the two-mode intensity difference noise heatmaps for the weak seed and strong seed cases for a propagation length $L=10$ cm. The correlation patterns are qualitatively different for the two cases. In each case, strong intercomb and intracomb correlations span the entire multi-pulse system. Certain pulses feature short-scale sharp variations in the correlation. This has been observed in ultrafast pulse propagation in nonlinear fibers and may be at least partially due to strong sensitivity of the (Kerr) nonlinear phase shift at high peak powers \cite{uddin2025probing}. In the present system, both self- and cross-phase modulation effects are significant given the megawatt level peak powers and can certainly affect the noise correlations we observe. We observe that a stronger seed appears to spread correlations over a larger fraction of each pulse's spectrum.

Finally, in Fig. \ref{fig:32} we explore the build-up of quantum correlations in both the time and frequency domain as the nonlinear interaction length increases. Time bin entanglement has been demonstrated in femtosecond pulses \cite{marcikic2002time} and we show here how it can be generalized to a multi-pulse system spanning a very broad spectral range. In particular, notice the strong intercomb correlations at long interaction length ($L=6,10$ cm) that are smeared out over several pixels. This indicates large parts of both pulses are strongly quantum correlated with one another, offering an enticing resource for quantum-enhanced pump-probe measurements, broadband entangled qubit generation for quantum communication, and more. This behavior extends to pulse pairs that do not couple directly through the three-wave mixing intrinsic to the system.

\section{Discussion}

In this work, we have presented a method of generating strong broadband quantum correlations and entanglement through the nonlinear interaction between multiple frequency combs. Furthermore, we have introduced tools that enable control over the covariance matrix in this highly multimode system, exemplifying their use in generating programmable two-mode squeezing of frequency modes in vastly different spectral ranges. We believe that this system holds exciting potential for quantum information multiplexing, quantum-enhanced spectroscopy, and more.

We have proposed experimental platforms for realizing the effects described here with both quasi-continuous wave and ultrafast sources. For true steady state multi-frequency comb operation, the latter should take the form of a synchronously pumped multimode cavity. As we describe in the SI, the pump powers and quality factors assumed in our simulations are typical of high $Q$ free space cavities pumped by commercially available $Q$-switched or mode-locked lasers \cite{shibuya2008terahertz, mackevivciute2025comparative}. Of course, the ultimate goal would be to shrink the form factor of such a device to monolithic or chip scale. To this end, exciting strides have been made to realize integrated and monolithic frequency combs and femtosecond optical parametric oscillators \cite{griffith2015silicon, hu2022high, jornod2023monolithically, gaeta2019photonic, chang2022integrated}. Although our focus on this work has been quantum correlations between frequency modes, as this is most relevant for applications in spectroscopy, we point to exciting proposals for using self-reconfiguring optics for identifying strongly squeezed supermodes in multimode nonlinear systems \cite{karnieli2025variational}. Amounting to a linear basis transformation, these proposals can readily be implemented using integrated Mach-Zender interferometer arrays. We envision exciting intersections of the physics explored here with these variational processors of quantum light.

Our mechanism realizes coupled frequency combs whose intercomb spacing $\omega_T$ far exceeds the intracomb spacing $\omega_m$. In practice, $\omega_m$ may lie at MHz or GHz frequencies, while $\omega_T$ can lie at microwave, terahertz, or infrared frequencies. Here, we have focused on the last regime, as it enables us to generate combs in highly disparate spectral ranges, but we emphasize that the physical mechanism holds irrespective of $\omega_m$ and $\omega_T$. This therefore admits a wide range of physical platforms (from optomechanical and microwave resonators to optical comb systems) to realize the physics described here.

We envision that our platform can also be used to explore other exciting physics beyond the quantum optics studied here. For example, topological band structures and non-Hermitian topology have been studied in synthetic frequency dimensions \cite{sloan2025noise, dutt2022creating, wang2021generating, wang2021topological, song2020two, cheng2023multi, pang2024synthetic, yang2025controlling}. Our platform realizes two synthetic frequency dimensions (the inter- and intracomb dimensions). By introducing non-Hermitian amplitude modulation in the intracomb dimension, one may be able to realize photonic higher order topological insulators as well as chirally protected multimode quantum states of light such as cluster states spanning distinct frequency combs. Furthermore, the interplay between second and third order nonlinearities in the multi-comb system is an exciting avenue to explore the intersection of multimode entanglement and single mode intensity noise squeezing \cite{rivera2023creating, pontula2025strong}. 

In conclusion, we have explored the quantum correlations and entanglement in a multi-frequency comb system generated by cascaded three-wave mixing mediated by a single idler frequency comb. This system holds promise for both fundamental studies of quantum optics in multiple dimensions as well as applications requiring on-demand generation of multimode quantum light across the electromagnetic spectrum.

\section{Acknowledgements} S.P. acknowledges the financial support of the Hertz Fellowship Program and NSF Graduate Research Fellowship Program. Y.S. acknowledges support from the University of Central Florida Office of Research through an internal research grant AWD00006722. M.S.\ acknowledges support from the U.S.\ Office of Naval Research (ONR) Multidisciplinary University Research Initiative (MURI) under Grant No.\ N00014-20-1-2325 on Robust Photonic Materials with Higher-Order Topological Protection. This material is based upon work supported in part by the Air Force Office of Scientific Research under the award number FA9550-20-1-0115; the work is also supported in part by the U. S. Army Research Office through the Institute for Soldier Nanotechnologies at MIT, under Collaborative Agreement Number W911NF-23-2-0121. This work is additionally supported in part by the DARPA Agreement No. HO0011249049. We also acknowledge support of Parviz Tayebati.

\bibliography{main}


\end{document}